\documentstyle[aps,epsf,prl,multicol]{revtex}

\begin{document}
\draft
\title{
Quantum chaos of  a kicked particle in a 1D infinite 
square potential well
} 
\preprint{HKBU-CNS-9815}
\author{ Baowen Li$^{a}$, Jie  
Liu$^{a,b}$, Yan Gu$^{a,c}$, and Bambi Hu$^{a,d}$
} 
\address{$^{a}$Department of Physics and Centre for Nonlinear Studies, 
Hong Kong Baptist University, Hong Kong, China  \\
$^{b}$ Institute of Applied Physics and Computational Mathematics, 
P.O.Box 8009, Beijing 100088, China\\
$^{c}$Center for Fundamental Physics, University of Science and 
Technology of China, Hefei, China\\
$^{d}$ Department of Physics, University of Houston, Texas 77204}
\date{\today} 
\maketitle

\begin{abstract} 

We study quantum chaos in a non-KAM system, i.e. a kicked
particle in a one-dimensional  infinite square potential well.  
Within the perturbative regime the classical
phase space displays stochastic web structures, and the diffusion
coefficient $D$ in the regime increases with the perturbative strength $K$ 
giving a scaling $D \propto K^{2.5}$, and in the large $K$ regime 
$D$ goes as $K^2$. Quantum mechanically, we observe that 
the level spacing statistics of the quasi eigenenergies 
changes from Poisson to Wigner distribution as the kick strength 
increases. The 
quasi eigenstates show power-law localization in the small $K$ region, which 
become extended one at large $K$. 
Possible experimental realization of this model is also discussed.

\end{abstract} 

\begin{multicols}{2} 

In the study of quantum chaos, most works are concentrated on those 
quantum systems whose classical counterparts obey the
Kolmogorov-Arnold-Moser (KAM) theorem.  That is to say,  
changing the external or driven parameter, the invariant curves gradually
break up and local chaos becomes global chaos, and the
classical motion becomes diffusive. The widely studied
models are the kicked rotator model (KRM) \cite{CCFI79,KRM} and quantum
billiards\cite{Billiard}. In these models, an ostensible phenomenon is the
dynamical localization, namely, the quantum suppression of
classical diffusion. This phenomenon was first discovered numerically by
Casati {\it et al.}\cite{CCFI79} in the KRM, and later
on confirmed by several experiments such
as the Rydberg atom in
microwave field \cite{DLExp} and an atom moving in a modulated 
standing wave etc.\cite{KRMEXP}. This phenomenon has been found to be
generic not only in the kicked quantum systems but also in 
the conservative Hamiltonian systems such as the quantum
billiards\cite{DLbild}, the Wigner band random matrix model\cite{CCGI}, 
and a single ion confined in a Paul trap\cite{Schleich97} and so on.

However, in addition to the systems 
mentioned
above which have been studied extensively in the past two decades,
there exists another class of systems which are out of the KAM frame.
In these systems, the invariant
curves do not exist at all for any small external/driven parameters.
Compared with the KAM systems, much less is known about quantum
chaos in such systems. For instance, we have only limited knowledge 
of the kicked harmonic oscillator introduced by Zaslavsky {\it et
al.}\cite{KHO,Chernikov} to describe a charged particle moving in a magnetic
field, and under the disturbance of a wave packet. This model can  also be
used to describe a single ion trapped in a harmonic potential\cite{GCZ97}.
This system is a degenerate one and does not satisfy the KAM 
theorem.  The quasi eigenenergies, the quasi eigenstates, and the long
time diffusion of 
this model cannot be studied  by using nowadays computer
facilities\cite{Dima92}, because its phase space 
is unbounded and cannot be reduced to a cylinder, as in the
case of the KRM.

The purpose of this letter is two-fold: (1) to construct a
simple non-KAM  system which could be investigated numerically both 
classically and quantum mechanically; (2) to
study quantum chaos in such a system.  As we shall see,
in spite of its simplicity, our model shows
stochastic webs
in classical 
phase space  
which is  the essential property
of a non-KAM system.
Unlike that of the kicked harmonic oscillator, the quasi eigenenergies and 
quasi eigenstates of this model can be computed 
easily. Furthermore, like the KRM, our model might be realized 
experimentally.  The study of this model aims to enrich our 
understanding of quantum chaos in the non-KAM  quantum systems.

The model we are considering in this letter is a particle moving inside a
one-dimensional (1D) infinite square potential well, and under 
the influence of a kicked 
periodic external potential.  The difference of this model from the 
kicked harmonic oscillator  lies in its phase configuration. As mentioned 
before, the phase space of the kicked harmonic oscillator is unbounded 
both in momentum and in space, whereas it is bounded on a  
cylinder with flattened end in our model. 

The Hamiltonian of our system is
\begin{eqnarray}
H =\frac{p^2}{2} + V_0(q) + k \cos(q+\alpha)\sum_{n=-\infty}^{\infty}
\delta(t-nT),\nonumber\\
V_0(q) =\left\{ {
\infty,\quad \mbox{for}\,q=0,\mbox{and}\,\pi \atop
0,\qquad \mbox{elsewhere},}
\right.
\label{Ham1}
\end{eqnarray}
$\alpha$ is a phase shift, in general case $\alpha \ne 0$. 
It is readily seen from this Hamiltonian that our model is a 
modification of 
the KRM. There are two minor changes: (1) two
hard walls are set up at $q=0$ and $\pi$, respectively; (2) a phase 
shift $\alpha$ for potential is made. The two hard walls destroy the 
analyticity of the potential, and make the model out of the KAM system.
The phase shift destroys the parity symmetry.

{\bf Classical Dynamics} -- The main characteristic of this  
system is the stochastic webs in the classical
phase space. Thus the diffusion can take place along the stochastic webs for
any small perturbation $K(=kT)$, see e.g. Fig. 1 for $K=0.01$, 
$\alpha=1$. 
This is the fundamental difference from the KAM system e.g. the KRM. In the 
KRM, for any $K < K_c=0.971635...$, 
no global classical diffusion occurs due to the invariant curves.
In this letter, we restrict our calculations to $\alpha=1$. It has been 
checked
that changing $\alpha$ does not change any results quantitatively, it 
just shifts the stochastic webs in phase space left or right.
The properties of the stochastic webs such as the thickness and symmetry 
etc. are also of great interest\cite{Chernikov}. We leave this part work 
over for the future study.
As the $K$ increases,  the stochastic layer becomes wider and wider, 
and eventually covers the whole phase space. In calculating
the diffusion coefficient for a given $K$, we have taken 10,000 points starting
from stochastic regions, and all the initial trajectories evolve for one 
million periods. Averages are taken over 10,000 trajectories for each time
period. It is found that the energy diffusion is asymptotically linear for
all values of $K$. 

The diffusion coefficient $D(\equiv \langle E_n \rangle/n)$ ($n$ is the time
in unit of $T$) versus $K$ is plotted in Fig. 2. It is evident
that there exist two different diffusion regions. For $K \gg 1$ the
diffusion coefficient is $D \sim K^2$, whereas for $ K\ll 1$, the 
diffusion behavior is
$D \sim K^{2.5}$ which is similar to that of the discontinuous twist
map\cite{TDM}. However, the underlying  mechanism is different.
In the case of the discontinuous twist map, the super slow 
diffusion is caused by the 
stickiness to the  cantori, whereas in our model it is due to the stable 
islands. 

Now we turn to quantum behaviors of this model.
One may ask: how do these kinds of characteristics, the stochastic 
webs and
the super slow diffusion, manifest themselves in  quantum mechanics in
terms of
the statistics of quasi eigenenergies and quasi eigenstates?
These are most interesting problems in the study of quantum 
chaos. Since our model is a 
periodically driven system, the evolution operator
over the period $T$ 
of the kick
is given by 
\begin{equation}
\hat U(T) = \exp(-\frac{i{\hat p}^2T}{4\hbar})
\exp(-\frac{iV({\hat q})}{\hbar})
\exp(-\frac{i{\hat p}^2T}{4\hbar}),
\label{Unit}
\end{equation}
where $V(q) = k \cos(q+\alpha)$.
The operator $\hat U(T)$
is also called the Floquet operator, it 
is time-reversal invariant. Moreover, it 
is unitary and satisfies the following 
eigenvalue equation, $\hat U(T)|\Psi_{\lambda}\rangle = 
e^{-\frac{i\lambda}{\hbar}}|\Psi_{\lambda}\rangle 
$
where the eigenphase $\lambda$ is real, $\lambda/T$ is the so called
quasi eigenenergy, and $\Psi_\lambda$ is  the quasi eigenstate (or the 
Floquet
 state).

The quasi eigenenergies 
can be obtained by diagonalizing $\hat U(T)$ within a large
number of  bases $|n\rangle$, which we  chose as the eigenstates of the
non-perturbative Hamiltonian system:
\begin{equation}
\langle q|n\rangle =\sqrt{\frac 2 \pi}\sin(nq),\,\, q \in [0,\pi];\,\,  
n=1,~2,...,~N. 
\label{basis}
\end{equation}
In our calculations $N$ is kept at 1024. (The calculation is also
performed with 512 
bases, but no quantitative difference is found.) The elements of 
matrix  are $U_{nm} = \langle n|\hat U(T)|m \rangle
$. As $\hat U(T)$ is a unitary operator, we construct
$\hat C = (\hat U(T) + \hat U^{+}(T))/2$ as a Hermitian operator. Then 
the elements of $C_{nm}= A_{nm} + i B_{nm}$. The matrix ${\bf A}$ and ${\bf 
B}$ satisfy the condition ${\bf A}^{T}={\bf A}$ and ${\bf B}^{T}=-{\bf B}$, 
respectively. $\left[\begin{array}{cc} {\bf A}&-{\bf B}\\
{\bf B}&{\bf A}
\end{array}\right]$ is  a $2N\times 2N$
 symmetric matrix. The standard algorithm\cite{NUMREC} is used to
diagonalize the above matrix and to obtain eigenvalues and
eigenvectors $\left[u, v\right]$.  Then we project the $N$ dimensional 
vector $u+iv$ 
on the basis of a plane wave to obtain the eigenstates of $\hat U(T)$.
The fast Fourier transform (FFT) of sinusoidal form\cite{NUMREC}
is employed to transform the wave function between the position
representation and energy representation in our calculations. 

{\bf Quasi eigenstates}-- The quasi eigenstates show
behaviors  quite different 
from that of the KRM. In the KRM, the
quasi eigenstates are exponentially localized in the momentum 
space\cite{KRM}. In
our model, however, the quasi-eigenstates are power-law localized, as is
shown in Fig. 3. In this figure, we demonstrate a few typical states at 
different values
of $K$ for localized, intermediate and extended ones. It is clearly seen 
that the localized states gradually transit to  extended ones as we 
increase the strength of  the kicked potential. This transition, as we shall 
see later, will manifest itself in the statistics of quasi eigenenergies. 

In fact, the power-law localization of the eigenstates can be traced back
to the structure of the matrix $U$.  In the KRM, the values of matrix
elements $U_{mm+n}$ decay faster than exponential when $n$ exceeds the
band width $b$ which is proportional to $K$, thus the elements outside this
band can be regarded as zero. Within the band of width $b$, the elements
are proved to be pseudorandom\cite{KRM}.  Such kind of band random matrix
attracted much attention in the past years. However, in our model the
situation is different. Careful analysis yields that the elements
outside the band decay as a power law with $|U_{m,m+n}| \approx 1/n^2$. 
We calculate $\langle U^2\rangle_n ( \equiv \langle
U^2_{mm+n}\rangle$) (the average is done over $m$) for four different $K$'s 
, and plot them in Fig. 4. 
The typical slope of the curves over a large range is approximately
minus $4$. And the band width $b$ in our model is found also to be
approximately proportional to the perturbation strength $K$. Inside this
band the magnitudes of the matrix elements are almost a constant. This
kind of band random matrix, describes a new class of
physical systems e.g. systems with non-analytic singular boundary,
has not yet been fully studied \cite{FM96}. More works are expect to be
done. 

{\bf Statistics of the quasi eigenenergies} -- 
The structure of the quasi eigenstates determines the energy level 
statistics. As is well known that the level repulsion can occur between the 
Floquet eigenvalues when the Floquet eigenfunctions
overlap. In the KRM, the quasi eigenfunctions are 
exponentially localized in angular momentum. Since the angular momentum  
has a finite range, the Floquet states with very close eigenvalues may lie 
so far
apart that they don't overlap. Thus, we don't have any level 
repulsion for these two eigenvalues. This is the reason why Poisson-like
spectral statistics persists in the KRM even though the 
system is classically
chaotic. To observe the transition from Poisson to Wigner distribution, one has
to consider a KRM defined on a torus\cite{KRM}.

In our model, however, the Floquet states do overlap in momentum 
space as is shown above. Therefore, we are expecting to observe the 
transition of the quasi eigenenergies statistics. 
The level spacing statistics of the quasi eigenenergies are shown in
Fig. 5 for four different values of $K=$ 0.1, 5, 25, and
50. This figure demonstrates a smooth transition from Poisson to 
Wigner distribution. To quantify this transition, the Brody 
distribution\cite{Brody} is used to 
best fit the above four distributions. (In fact we use the 
cumulative distribution function $I(s) = \int_0^{s} P(s')ds'$). The 
best fitting gives 
rise to the Brody parameter $\beta=0.03, 0.08, 0.46$ and $0.82$, 
for the four distributions in Fig. 5, respectively. To check the approach
to Poisson distribution as $K$ goes down to zero, the $P(s)$ is also
calculated for
$K=10^{-4}$, as expected, we obtain a good Poisson distribution, the
best fitting gives rise to $\beta =0.01$.

It must be stressed that the difference between our model and the KRM
shown in level spacing statistics of the quasi eigenenergies and the 
quasi eigenstates comes from non-analyticity of the potential, which makes
the phase space in our model a half cylinder with the end flattened.
This non-analyticity also leads to a different structure of the evolution 
matrix $U$. Moreover, we would like to point out that since  the KRM  
has already been realized in laboratory by putting the cold (sodium/cesium) 
atoms 
in a periodically pulsed standing wave of light\cite{KRMEXP}, our model 
could be also realized experimentally. One possible way is to put the 
cold atoms in a quasi 1D quantum dot. The atoms are then driven by a 
periodically pulsed 
standing wave of light. The quasi 1D quantum 
dot might be realized by formulating a 2D quantum dot elongated in one 
direction, namely, the size in 
one direction is much larger than another. This experiment would allow 
us to study quantum chaos in a non KAM system.

In summary, we have studied the classical dynamics and quantum behaviors 
of a kicked particle in a 1D infinite square potential well. In spite of its 
simplicity, our
model exhibits stochastic webs which is one of  the basic features of 
the non-KAM systems. The classical dynamics is diffusive for any 
infinitesimal perturbation, and the 
classical diffusion rate is found to be $D \propto K^{2.5}$ for small
values of 
$K$, and $D \propto K^2$ for large values of $K$. The level 
statistics of quasi eigenenergies shows a smooth transition  from Poisson to 
Wigner distribution for a fixed dimension of the Floquet matrix. The quasi eigenstates
are found to be power-law localized with exponent equal to two.
Our model provides a new 
paradigm in investigating classical quantum
correspondence of stochastic motion exhibited in Hamiltonian systems with
non-analytic boundary conditions.
\\

BL would like to thank F. Borgonovi, G. Casati, Y. Fyodorov, F. 
Izrailev, and A. Mirlin for helpful discussions. He also thanks the Abdus 
Salam International Centre for Theoretical Physics, Trieste, Italy, for 
the kind hospitality, where this paper was finished. We are grateful to
the referees for useful suggestions and comments.  The work
was supported in part by grants from the Hong Kong Research Grants Council
(RGC) and the Hong Kong Baptist University Faculty Research Grant (FRG).


\begin{figure}
\narrowtext
\caption
{
A typical classical phase space of our model at very small perturbation
strength. The stochastic webs is clearly seen. Here we have $K=0.01$,
$\alpha=1$. One trajectory starts from $(q_0=0.1,p_0=0.012)$ and evolves
for 100,000 periods. 
}
\end{figure}

\begin{figure}
\epsfxsize=8cm
\epsfbox{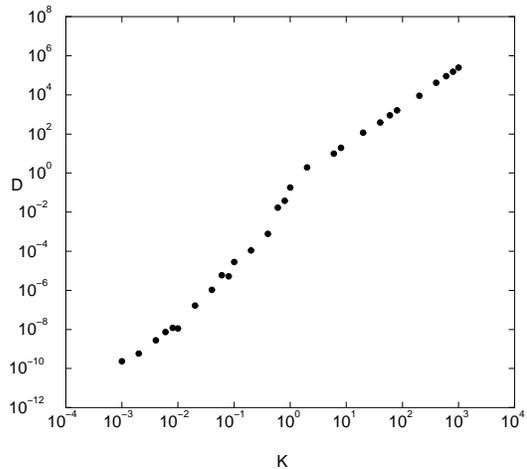}
\vspace{-2.5cm}
\narrowtext
\caption
{
Classical diffusion coefficient $D$ versus perturbation strength $K$. 
The best fitting by using the data $K >1$ gives rise to a slope 1.97, 
whereas that by using the data $K<0.1$ gives rise to a slope 2.47. A 
clear turning point  of the slope can be seen at $K$ about 1.} 
\end{figure}

\begin{figure}
\epsfxsize=8cm
\epsfbox{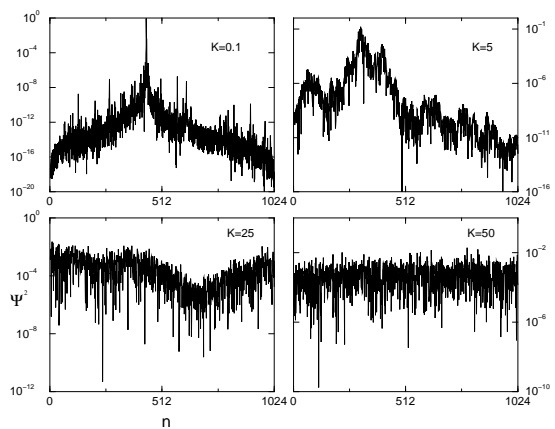}
\vspace{-2.5cm}
\narrowtext
\caption
{Typical quasi eigenstates in different regimes, 
localized ($K=0.1$ and $5$), intermediate ($K=25$), and extended 
($K=50$). The corresponding values of $K$ are shown in the figure.
}
\end{figure}

\begin{figure}
\epsfxsize=8cm
\epsfbox{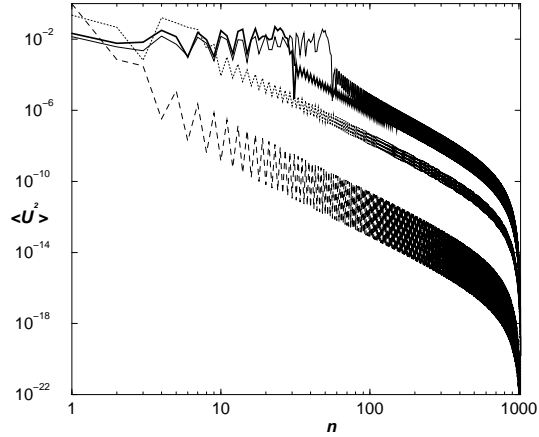}
\vspace{-2.5cm}
\narrowtext
\caption
{The averaged matrix element $\langle U^2\rangle$ versus $n$ for different 
values of $K$.
See the text for its definition. From left to 
right, the dashed curve for $K=0.1$, dotted curve for 5, thick solid 
curve for 25, and thin solid curve for 50. The band width is about the 
order of $K$, which is clearly seen from the figure. The slopes of these 
four curves are about minus four.} 
\end{figure}

\begin{figure}
\epsfxsize=8cm
\epsfbox{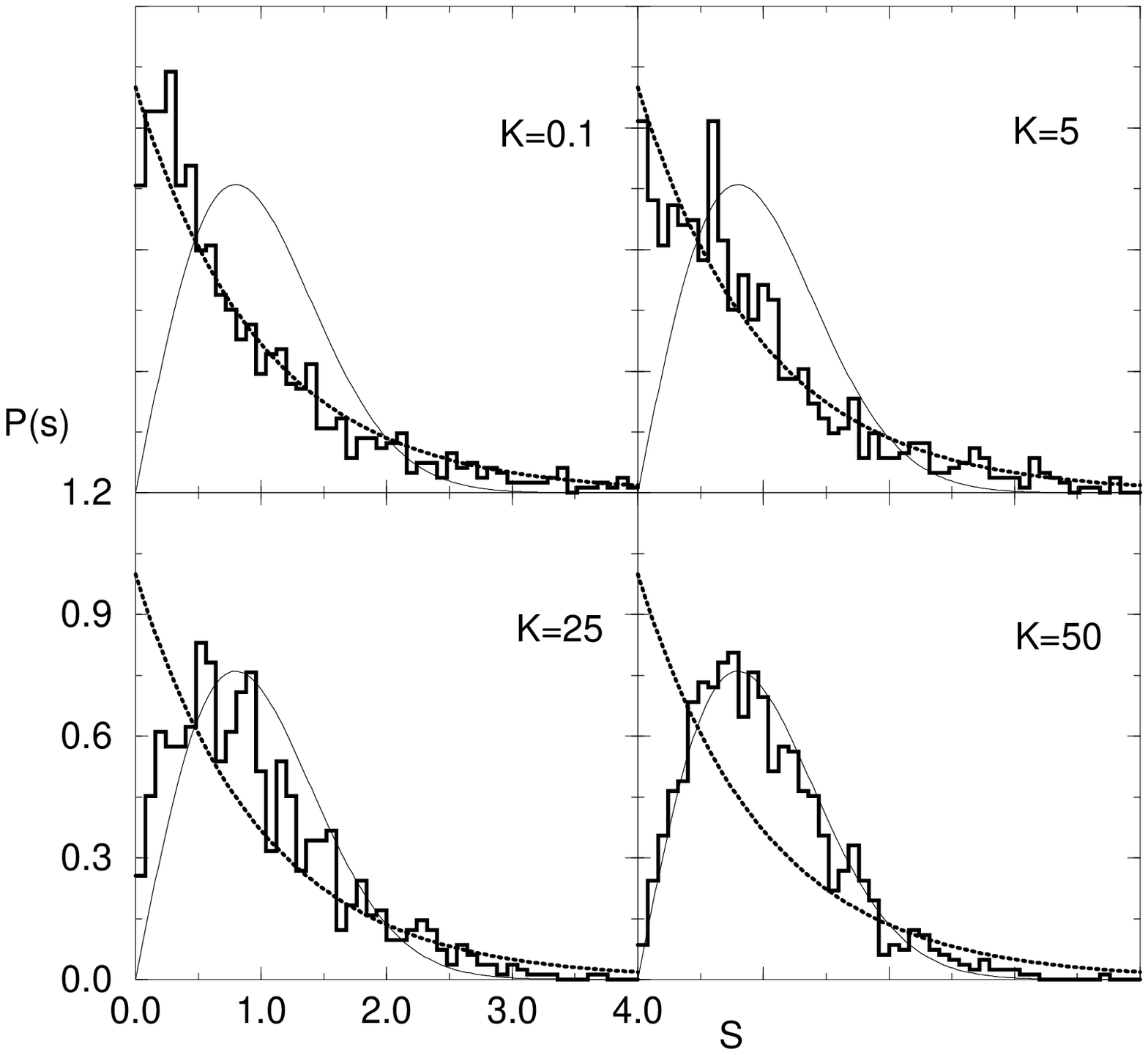}
\vspace{-2.cm}
\narrowtext
\caption{
The distribution of the nearest neighbor level spacing $P(s)$ for 
different $K$. The corresponding values of $K$ are given in the 
figures. The 
dotted curve is Poisson and the thin curve is Wigner distribution. The 
histograms are numerical results. $P(s)$ at $K=0.1$ and 5 are     
close to  Poisson distribution and at $K=25$ is  intermediate, and $K=50$ is 
close to Wigner distribution. The corresponding best fitting Brody
parameter $\beta$ are 
$0.03, 0.08, 0.46$, and $0.83$, respectively.}
\end{figure}

\end{multicols}

\begin{references}
\bibitem{CCFI79}G.~Casati, B.~V.~Chirikov, J.~ Ford, and F.~M.~Izrailev, 
Lect. Notes Phys. {\bf 93}, 334 (1979); see also [2].

\bibitem{KRM}
B.~V.~Chirikov, Phys. Rep. {\bf 
52}, 263 (1979); F.~M.~Izrailev, Phys. Rep. {\bf 196}, 299 (1990). G. Casati 
and B.~V.~Chirikov, {\it Quantum Chaos} (Cambridge University Press, 
Cambridge, 1995).

\bibitem{Billiard}M.~J.~Giannoni,
A.~Voros, and J.~Zinn-Justin" (eds) {\it Chaos and Quantum Systems}
(Amsterdam, Elsevier 
1991); J. Stat. Phys.{\bf 83}, (1995) (a special issue devoted to 
quantum billiards).

\bibitem{DLExp}E.~J.~Galvez, B. E. Sauer, L. Moorman, P. M. Koch, and D. 
Richards, Phys. Rev. Lett. {\bf 61}, 2011 (1988); 
J.~E.~Bayfield, G.~Casati, I.~Guarneri, and D.~Sokol, {\it ibid.} {\bf
63}, 364
(1989); M. Arndt, A. Buchleitner, R.~N.~Mantegna, and H.~Walther {\it
ibid.} 
{\bf 67}, 2435 (1991).

\bibitem{KRMEXP}
F. L. Moore, J. C. Robinson, C. F. Bharucha, P. E. Williams, and M. G. 
Raizen, Phys. Rev. 
Lett. {\bf 73}, 2974 (1994); J. C. Robinson, C. F. Bharucha, F. L Moore, 
R. Jahnke, G. A. 
Georgakis, Q. Niu, M. G. Raizen, and B. Sundaram, {\it ibid.} {\bf 74}, 3963 
(1995);
 F.~L.~Moore, J. C. Robinson, C. F. Bharucha, B. Sundaram, and M. G. Raizen
{\it ibid.} {\bf 75}, 4598 (1995);
B. G. Klappauf, W. H. Oskay, D. A. Steck, and M. G. Raizen, {\it ibid.} {\bf 
81}, 4044 (1998).

\bibitem{DLbild}F.~Borgonovi, G.~Casati, and, B.~Li, Phys. Rev.
Lett. {\bf 77} 4744 (1996); K.~M.~Frahm and D.~L.~Shepelyansky, {\it
ibid.} {\bf 78}, 1440 (1997).

\bibitem{CCGI}G.~Casati, B.~V.~Chirikov, I.~Guarneri, and F.~M.~Izrailev,
Phys. Rev.
E {\bf 48}, R1613 (1993), G.~Casati, B.~V.~Chirikov, I.~Guarneri, and 
F.~M.~Izrailev, Phys. Lett. A {\bf 223}, 430(1996).

\bibitem{Schleich97}M.~El~Ghafar, P. T\"orm\"a, V. Savichev, E. Mayr, A. 
Zeiler, and W. P. Schleich, Phys. Rev. Lett. {\bf 78}, 4181 (1997).

\bibitem{KHO} 
G.~M.~Zaslavsky, R.~Z.~Sagdeev, D.~A.~Usikov and 
A.~A.~Chernikov, "{\em Weak Chaos and Quasi-regular Patterns}", and the 
references therein, Cambridge University Press, 1992;
G.~P.~Berman, 
V.~Yu.~Rubaev and G.~M.~Zaslavsky, Nonlinearity {\bf 4}, 543 (1991).

\bibitem{Chernikov} A. A. Chernikov, R. 
Z. Sagdeev, D. A. Usikov, M. Yu. Zakharov, and G. M. Zaslavsky, 
Nature {\bf 326}, 559 (1987); A. A. Chernikov, R. Z. Sagdeev 
and G. M. Zaslavsky, Physica D {\bf 33}, 65 (1988).

\bibitem{GCZ97} S.~A.~Gardiner, J.~I.~Cirac, and P.~Zoller, Phys. Rev.
Lett. {\bf 79}, 4790 (1997).

\bibitem{Dima92} D. Shepelyansky and C. Sire. Europhys. 
Lett. {\bf 20}, 95 (1992); F. Borgonovi and L. Rebuzzini, Phys. Rev. E
{\bf 52}, 2302 (1995); B. Hu, B. Li, J. Liu, and J.-L Zhou, {\it ibid.} 
{\bf 58}, 1743 (1998).


\bibitem{TDM}F.~Borgonovi, Phys. Rev. Lett. {\bf 80}, 4653 (1998); 
 F.~Borgonovi, P. Conti, D. Rebuzzi, B. Hu, and B. Li, 
Physica D, (1998) (in press).

\bibitem{NUMREC}W. H. Press, S. A. Teukolsky, W. T. Vetterling, and B. P. 
Flannery, {\it Numerical Recipes in Fortran}, Cambridge University Press, 
1992.

\bibitem{FM96}A. M. Mirlin, Y. V. Fyodorov, F. M. Dittes, J. Quezada and 
T. H. Seligman, Phys. Rev. E {\bf 54}, 3221 (1996).

\bibitem{Brody}T. A. Brody, J. Flores, J. B. French, P. A. Mello, A. 
Pandey and S. S. M. Wong, Rev. Mod. Phys. {\bf 53}, 385 (1981).


\end{references}
\end{document}